\documentstyle[11pt,axodraw,colordvi]{article}
\textheight
22.0cm \textwidth 15cm
\begin{document}
\begin{center}
\vspace{.5cm}{\large\bf Bloch space structure of cascade, lambda and vee type \\ of three-level systems and qutrit wave function}
\end{center}
\begin{center}
\vspace{2cm}{\bf Surajit Sen,{\footnote[1]{ssen55@yahoo.com}} Mihir
Ranjan Nath{\footnote[2] {mrnath\_95@rediffmail.com}} and Tushar
Kanti Dey {\footnote[3] {tkdey54@rediffmail.com}}
\\Department of Physics\\
Guru Charan College\\ Silchar 788004, India \\
\vspace{.5cm} Gautam Gangopadhyay
{\footnote[4]{gautam@bose.res.in}}\\
S N Bose National Centre for Basic Sciences\\
JD Block, Sector III
\\Salt Lake City, Kolkata 700098, India}\\
\vspace{1cm}
\end{center}
\begin{center}
\hrule width 5.7in height .25pt depth .5pt
\end{center}
\begin{abstract}
The cascade, lambda and vee type of three-level systems are shown to be described by three different Hamiltonians in the $SU(3)$ basis. We investigate the Bloch space structure of each configuration by solving the corresponding Bloch equation and show that at resonance, the seven-dimensional Bloch sphere ${\mathcal S}^7$ is broken into two distinct subspaces ${\mathcal S}^2{\times}{\mathcal S}^4$ due to the existence of a pair of quadratic constants. We also give a possible representation of the qutrit wave function and discuss its equivalence with the three-level system.
\end{abstract}
\vspace{.175cm}
\begin{center}
\hrule width 5.7in height .25pt depth .5pt
\end{center}
\vspace{2cm}
\par
{\bf PACS No.: 42.50.Ct, 42.50.Pq, 42.50.Ex}
\par
{\bf Key words: $SU(3)$ group; Bloch equation; Three-level system; Qutrit}
\vfill
\pagebreak
\par
The atomic system interacting with two or multiple number of lasers provides an unique opportunity to manipulate them because of its well-defined level structure. To develop a systematic theory of the coherent control mechanism, it is necessary to study the dynamics of the two-, three- or multi-level systems which may pave the way to the experimental realization of the quantum computer [1-3] and associated phenomena such as quantum teleportation [4], quantum cryptography [5], dense coding [6,7], quantum cloning [8] etc. In the quantum information theory parlance where the qubit is treated as the primary building block, the Bloch space representation of the two-level system is significant for many reasons. Most importantly, a qubit state $|q_B>$ is codified as the superposition of the computational basis \{$|0>$, $|1>$\} and it corresponds to various points on the two-dimensional unit Bloch sphere ${\mathcal S}^2$ [3,9]. Thus a quantum mechanical gate, which indeed represents an unitary operator $A$, corresponds to the transformation of the qubit state as $|q_B'>=A|q_B>$, defined on the surface of that sphere. A natural extension of the qubit is the qutrit system, where the computational basis are defined as \{$|0>$, $|1>$ and $|2>$\}, respectively. Physically a qutrit possesses much complex but richer structure than the ordinary qubit and there exists several suggestions that it may be implemented either as a three-level system [10,11], or transverse spatial modes of single photons [12], or polarization states of the biphoton field [13,14], respectively. In recent years considerable progress has been made to understand various aspects of the qutrit system treating it a three-level system  driven by bichromatic laser fields. These studies include, the quantum information transfer between qutrits [15], separability among joint states of qutrits [16], entanglement sudden death for qubit-qutrit composite system [17], cooperative mode in the qutrit system with complex dipole-dipole interaction [18], teleportation between two unknown entangled qutrits [19], cloning the qutrit [20], possible algebraic and geometric structure [21-24], new quantum key distribution protocols dealt with qutrits [25,26] etc [27]. In spite of these progress, understanding the Bloch space of all three-level configurations and the wave function of the qutrit system is the necessary prerequisite to define various logic gates and the associated circuits, which perhaps form the basis of developing a richer form of quantum information theory beyond qubit.
\par
A systematic study of the N-level system and its connection with the $SU(N)$ group was first initiated by Eberly and Hioe [28-30] and later it was investigated in detail in context with the three-level system [31-38]. From these studies it is revealed that the quadratic casimir for the $SU(3)$ group is manifested through the existence of a set of nonlinear constants which gives rise to the non-trivial structure of the Bloch space, i.e., ${\mathcal S}^0{\times}{\mathcal S}^2{\times}{\mathcal S}^4$. Later these constants were studied by solving the pseudo-spin equation [35] and also by the Floquet theory technique [37,38]. However, all three-level configurations, namely, cascade, lambda and vee type of system, are found to exhibit same set of nonlinear constants. The primary hindrance of getting different constants for different configurations is the following: It was pointed out by Hioe and Eberly that if the position of the intermediary energy level $E_2$ is changed with respect to other two levels shown in Fig.1, then all three-level configurations can be expressed by similar Hamiltonian  [29]. Since the proposal was made, it has become a widely accepted convention that the cascade, lambda and vee type of three-level systems are described by similar Hamiltonian with the energy hierarchy conditions, namely, $E_3>E_2>E_1$, $E_2>E_3>E_1$ and $E_1>E_2>E_3$, respectively. Similar structure of the Hamiltonian leads to same Bloch space for three systems and therefore forbids the existence of different non-linear constants for different configurations. However, it is now well-known that each configuration of the three-level system is associated with a diverse range of
coherent phenomena and therefore must be described by the distinct Hamiltonian. For example, the lambda configuration exhibits the phenomena such as STIRAP [39], EIT [40] etc, while the vee configuration is associated with quantum jump [41], quantum zeno effect [42], quantum beat [43] and others, respectively. Taking the interacting fields to be either classical or quantized bichromatic fields, the necessity of the inequivalence between the lambda and vee systems is also studied [44]. In a recent work, we have proposed a novel scheme to write the Hamiltonians of the three-level systems in the $SU(3)$ operator basis where the energy levels are arranged as $E_3^a>E_2^a>E_1^a$ ($a=\Xi, \Lambda$ and $V$) shown in Fig.2 [45].
In particular, we have discussed the exact solution of the semiclassical and quantized three-level systems and have compared the corresponding Rabi oscillations.
\par
The primary objective of this paper is to discuss the structure of the Bloch space of the three-level systems, while taking distinct Hamiltonian for different configuration. It is shown that at vanishing detuning frequency, the Bloch space of each three-level system is broken into two distinct sectors ${\mathcal S}^2{\times}{\mathcal S}^4$ and they are different for different configurations. Finally we have constructed the wave function of the qutrit system and discuss its various properties.
\par
It is customary to view the three-level system as two two-level systems with one of their levels overlaps with each other. Thus in the dipole approximation, the Hamiltonian of the semiclassical cascade system can be expressed as [45]
$${\rm H^{\Xi}} = \hbar(\omega_1U_3+\omega_2T_3)+{\bf \hat{d}_{12}}\cdot {\bf E_1}(\Omega_1)+{\bf \hat{d}_{23}}\cdot {\bf E_2}(\Omega_2),\eqno (1)$$
where, $\bf{\hat{d}_{12}}=(U_{+}+U_{-})$  and $\bf{\hat{d}_{23}}=(T_{+}+T_{-})$ be the dipole operators representing the transitions $1\leftrightarrow 2$ and $2\leftrightarrow 3$ and
${\bf E_1}(\Omega_1)={\bf E}_{1+}+{\bf E}_{1-}$ and ${\bf E_2}(\Omega_2)={\bf E}_{2+}+{\bf E}_{2-}$ be the bichromatic classical electric fields with frequencies $\Omega_1$ and $\Omega_2$, respectively. Similarly we have
$${\rm H^{\Lambda}} = \hbar(\omega_1V_3+\omega_2T_3)+{\bf \hat{d}_{13}}\cdot {\bf E_1}(\Omega_1)+{\bf \hat{d}_{23}}\cdot {\bf E_2}(\Omega_2),\eqno (2)$$
for the lambda system, where $\bf{\hat{d}_{13}}=(V_{+}+V_{-})$, $\bf{\hat{d}_{23}}=(T_{+}+T_{-})$ be the dipole operators representing transitions $1\leftrightarrow 3$ and $2\leftrightarrow 3$, respectively. Finally for the
vee system we have
$${\rm H^{V}} = \hbar(\omega_1V_3+\omega_2U_3)+{\bf \hat{d}_{13}}\cdot {\bf E_1}(\Omega_1)+{\bf \hat{d}_{12}}\cdot {\bf E_2}(\Omega_2),\eqno (3)$$
where $\bf{\hat{d}_{13}}=(V_{+}+V_{-})$ and $\bf{\hat{d}_{12}}=(U_{+}+U_{-})$ characterize the transitions $1\leftrightarrow 3$ and $1\leftrightarrow 2$, respectively. In Eq.(1-3), $\hbar \omega _1 (=-{E}_1^\Xi), \hbar (\omega _1-\omega_2) (= { E}_2^\Xi)$ and $\hbar\omega _2 ({=E}_3^\Xi)$ be the energies of the three levels of the cascade system, $\hbar \omega _1 (=-{E}_1^\Lambda), \hbar \omega _2 (= -{ E}_2^\Lambda)$ and $\hbar(\omega _2 + \omega _1 )({=E}_3^\Lambda)$ be the energies of the lambda system and $\hbar \omega _1 (={E}_1^V), \hbar \omega _2 (= { E}_2^V)$ and $-\hbar(\omega _2 + \omega _1 )({=E}_3^V)$ be those of the vee system shown in Fig.2, respectively. We note that in all cases, unlike the conventional scheme shown Fig.1 [29], the energy levels maintain the hierarchy, $E_3^a>E_2^a>E_1^a$ ($a=\Xi, \Lambda$ and $V$). In Eqs.(1-3), we have defined the $SU(3)$ shift vectors,
\begin{flushleft}
\hspace{0.5in}
 $T_{\pm}=\frac{1}{2}(\lambda_1 \pm i\lambda_2)$,
\hspace{0.35in}   $U_{\pm}=\frac{1}{2}(\lambda_{6} \pm i\lambda_{7})$,
\hspace{0.35in}   $V_{\pm}=\frac{1}{2}(\lambda_{4} \pm i \lambda_5)$,
\end{flushleft}
\begin{flushleft}
\hspace{0.5in}  $T_3=\lambda_3$,
\hspace{0.35in}  $U_3=(\sqrt{3}\lambda_8-\lambda_3)/2$,
\hspace{0.35in}  $V_3=(\sqrt{3}\lambda_8+\lambda_3)/2$,\hfill (4)
\end{flushleft}
respectively, where the Gellmann matrices are given by,

\begin{flushleft}
\hspace{0.5in}
 $\lambda_1   = \left[ {\begin{array}{*{20}c}
   0 & 1 & 0  \\
   1 & 0 & 0  \\
   0 & 0 & 0  \\
\end{array}} \right],
\hspace{0.35in} \lambda_ 2   = \left[ {\begin{array}{*{20}c}
   0 & -i & 0  \\
   i & 0 & 0  \\
   0 & 0 & 0  \\
\end{array}} \right],
\hspace{0.35in} \lambda_3   = \left[ {\begin{array}{*{20}c}
   1 & 0 & 0  \\
   0 & -1 & 0  \\
   0 & 0 & 0  \\
\end{array}} \right],$
\end{flushleft}

\begin{flushleft}
\hspace{0.5in}
 $\lambda_4   = \left[ {\begin{array}{*{20}c}
   0 & 0 & 1  \\
   0 & 0 & 0  \\
   1 & 0 & 0  \\
\end{array}} \right],
\hspace{0.35in} \lambda_5   = \left[ {\begin{array}{*{20}c}
   0 & 0 & -i  \\
   0& 0 & 0  \\
   i & 0 & 0  \\
\end{array}} \right],
\hspace{0.35in} \lambda_6   = \left[ {\begin{array}{*{20}c}
   0 & 0 & 0  \\
   0 & 0 & 1  \\
   0 & 1 & 0  \\
\end{array}} \right],$
\end{flushleft}

\begin{flushleft}
\hspace{0.5in}
 $\lambda_7   = \left[ {\begin{array}{*{20}c}
   0 & 0 & 0  \\
   0 & 0 & -i  \\
   0 & i & 0  \\
\end{array}} \right],
\hspace{0.25in} \lambda_8= \frac{1}{\sqrt{3}}\left[
{\begin{array}{*{20}c}
   1 & 0 & 0  \\
   0 & 1 & 0  \\
   0 & 0 & -2  \\
\end{array}} \right],\hfill (5)$
\end{flushleft}
respectively. These matrices follow the commutation and anti-commutation relations
\begin{flushleft} \hspace{1.5in}
$[\lambda_{i},\lambda_{j}]=2 i f_{ijk}\lambda_k$,
\hspace{0.35in}
$\{\lambda_{i},\lambda_{j}\}=\frac{4}{3}\delta_{ij}+2
d_{ijk}\lambda_k$,\hfill (6)
\end{flushleft}
where $d_{ijk}$ and $f_{ijk}$ ($i,j=1,2,..,8$) represent completely symmetric and completely antisymmetric structure constants, respectively, which characterize $SU(3)$ group [46].
\par
In the rotating wave approximation (RWA), the Hamiltonians can be written as
$$H^{\Xi} = \hbar(\Omega
_1-\omega_1-\omega_2) U_3+\hbar(\Omega _2-\omega_1-\omega_2) {\rm
T}_3+\hbar(\Delta^{\Lambda}_1U_3+\Delta^{\Lambda}_2T_3)+$$
$$\hbar \kappa _1 U_ + \exp ( - i\Omega _1 t) +\hbar \kappa _2 T_+\exp(-i\Omega_2 t)+h.c. \quad,\eqno (7)$$
for the cascade system,
$$H^{\Lambda} = \hbar(\Omega
_1-\omega_1-\omega_2) V_3+\hbar(\Omega _2-\omega_1-\omega_2) {\rm
T}_3+\hbar(\Delta^{\Lambda}_1V_3+\Delta^{\Lambda}_2T_3)+$$
$$\hbar \kappa _1 V_ + \exp ( - i\Omega _1 t) +\hbar \kappa _2 T_+\exp(-i\Omega_2 t)+h.c. \quad,\eqno (8)$$
for the lambda system,
$$H^{V} = \hbar(\Omega
_1-\omega_1-\omega_2) V_3+\hbar(\Omega _2-\omega_1-\omega_2) {\rm
U}_3+\hbar(\Delta^{\Lambda}_1V_3+\Delta^{\Lambda}_2U_3)+$$
$$\hbar \kappa _1 V_ + \exp ( - i\Omega _1 t) +\hbar \kappa _2 U_+\exp(-i\Omega_2 t)+h.c. \quad,\eqno (9)$$
for the vee system, respectively. In Eq.(7-9), $\kappa_{i}$ ($i=1,2$) be the coupling parameters
and $\Delta^{a}_1=(2\omega_1+\omega_2-\Omega_1)$ and $\Delta^{a}_2=(\omega_1+2\omega_2-\Omega_2)$ be the respective detuning frequencies. Thus we note that the Hamiltonian of any specific three-level configuration can be expressed by a subset of $SU(3)$ matrices, namely,
\begin{flushleft}
\hspace{0.5in} Cascade system : \hspace{0.3in} $U_{\pm}, U_3, T_{\pm}, T_3$, i.e., $\{\lambda_{1}, \lambda_{2}, \lambda_{3}, \lambda_{6}, \lambda_{7},
\lambda_{8}\}$,

\hspace{0.5in} Lambda system : \hspace{0.3in} $T_{\pm}, T_3, V_{\pm}, V_3$, i.e.,
$\{\lambda_{1}, \lambda_{2}, \lambda_{3}, \lambda_{4}, \lambda_{5},
\lambda_{8}\}$,               \hfill(10)

\hspace{0.5in} Vee system \hspace{0.32in}: \hspace{0.32in} $U_{\pm}, U_3, V_{\pm}, V_3$, i.e.,
$\{\lambda_{3}, \lambda_{4}, \lambda_{5}, \lambda_{6}, \lambda_{7},\lambda_{8}\}$,
\end{flushleft}
respectively [45], rather than all eight matrices [28-30]. Given with three well-defined models, we proceed to develop the Bloch equation of all three-level systems and discuss the constraints involving their solutions.
\par
Let the solution of the Schr\"{o}dinger equation of a generic three-level system described by the Hamiltonian Eq.(7-9) is given by,
$$ \Psi^{a} (t) = C_0^a(t) \left| 0 \right\rangle  + C_1^a(t)
\left| 1 \right\rangle  + C_2^a(t) \left| 2 \right\rangle,\eqno (11)$$
where $C_0^a(t)$, $C_1^a(t)$ and $C_2^a(t)$ be the normalized amplitudes with basis states $\left| 0 \right\rangle$, $\left| 1 \right\rangle$ and $\left| 2 \right\rangle$, respectively. The exact evaluation of the probability amplitudes enables us to calculate the density matrix of any system given by
$$\rho^{a}(t)=|\Psi^a(t)>\otimes<\Psi^a(t)|.\eqno (12)$$
To start with, we consider the dressed wave function of the cascade system obtained by a unitary transformation
$$ \tilde {\Psi}^{\Xi}=U_{\Xi}^\dag(t) {\Psi}^{\Xi},\eqno (13)$$
where the unitary operator is given by
$$ U_{\Xi}(t) = exp [-\frac{i}{3} ((\Omega _1 + 2\Omega
_2)T_3  + (2 \Omega _1 + \Omega _2 )U_3 )t].\eqno (14)$$
The corresponding time-independent Hamiltonian in Eq.(7) of the cascade system is given by
\begin{flushleft}\hspace{0.95in}
$\tilde {H^\Xi}(0) =
( - \hbar U_{\Xi}^ \dagger  \dot U_{\Xi} + U_{\Xi}^ \dagger  {H^\Xi(t)}U_{\Xi})$
\end{flushleft}
\begin{flushleft} \hspace{1.5in}$
=\left[ {\begin{array}{*{20}c}
   {\frac{1}{3}\hbar (\Delta_1^\Xi+2\Delta_2^\Xi) } & {\hbar\kappa_2  } & 0  \\
   {\hbar\kappa_2  } & {\frac{1}{3}\hbar (\Delta_1^\Xi-\Delta_2^\Xi) } & {\hbar\kappa_1  }  \\
   0 & {\hbar\kappa_1  } & { - \frac{1}{3}\hbar (2\Delta_1^\Xi+\Delta_2^\Xi) }   \\
\end{array}} \right].$\hfill(15)
\end{flushleft}
\par
Similarly, the unitary operator of the lambda system described by the Hamiltonian Eq.(8) is,
$$U_{\Lambda}(t) = exp[-\frac{i}{3} ((2\Omega _2 -
\Omega _1)T_3  + (2 \Omega _1 - \Omega _2 )V_3 )t],\eqno (13)$$
and the corresponding time-independent Hamiltonian is given by,
\begin{flushleft}
\hspace{1.0in}$ \tilde {\rm H^{\Lambda}}(0) = \left[ {\begin{array}{*{20}c}
   {\frac{1}{3}\hbar( \Delta_1^\Lambda+\Delta_2^\Lambda) } & {\hbar\kappa_2  } & {\hbar\kappa_1  }  \\
   {\hbar\kappa_2  } & {\frac{1}{3}\hbar (\Delta_1^\Lambda-2\Delta_2^\Lambda) } & 0  \\
   {\hbar\kappa_1  } & 0 & {  \frac{1}{3}\hbar (\Delta_2^\Lambda-2\Delta_1^\Lambda) }   \\
\end{array}} \right].$\hfill(17)
\end{flushleft}
\par
Also the Hamiltonian of the vee system in Eq.(6) can be made time-independent by using the transformation operator
\begin{flushleft}
\hspace{1.5in}$ U_{V}(t) = exp[-\frac{i}{3} ((2\Omega _2 - \Omega
_1)U_3  + (2 \Omega _1 - \Omega _2 )V_3 )t],$ \hfill(18)
\end{flushleft}
and corresponding Hamiltonian is,
\begin{flushleft}
\hspace{1.0in}$ \tilde {H^V}(0) = \left[ {\begin{array}{*{20}c}
   {\frac{1}{3}\hbar (2\Delta_1^V-\Delta_2^V) } & 0 & {\hbar\kappa_1  }  \\
   0 & {\frac{1}{3}\hbar (2\Delta_2^V-\Delta_1^V) } & {\hbar\kappa_2  }  \\
   {\hbar\kappa_1  } & {\hbar\kappa_2  } & { - \frac{1}{3}\hbar (\Delta_1^V+\Delta_2^V) }   \\
\end{array}} \right].$\hfill(19)
\end{flushleft}
Thus we have three distinct Hamiltonians with different non-vanishing entries for three different configurations.
\par
To obtain the Bloch equation, we define the generic $SU(3)$ Bloch vectors
$$S_{i}^a(t)=Tr[\rho^a(t)\lambda_i], \eqno (20)$$
where $\rho^a$ be the density matrix of the three-level systems given by Eq.(12) which satisfies the Lioville equation, namely,
$$\frac{d\rho^a}{dt}=\frac{i}{\hbar}[\rho^a, \tilde{H}^a(0)].\eqno (21).$$
From Eq.(20), the density matrix written in terms of Bloch vector is,
$$\rho^a(t)=\frac{1}{3}({\mathbf 1}+ \frac{3}{2}\sum\limits_{i = 1}^8 {S_i^a(t) \lambda_i } ).\eqno (22)$$
Substituting Eq.(22) and the Hamiltonians given by Eqs.(15), (17) and (19) in Eq.(21), we obtain the Bloch equation
$$\frac{dS_i^a}{dt}=M_{ij}^aS_j^a,\eqno (23)$$
where $M_{ij}^a$ be the eight dimensional anti-asymmetric matrix. For the Hamiltonian of the cascade system given by Eq.(15), the matrix $M_{ij}^\Xi$ reads
$$ M_{ij}^{\Xi}  =
\left[ {\begin{array}{*{20}c}
   0 & \Delta_2^{\Xi} & 0 & 0  & -\kappa_1 & 0 & 0 & 0  \\
   -\Delta_2^{\Xi} & 0 & 2\kappa_2 & \kappa_1 & 0 & 0 & 0 & 0  \\
   0 & -2 \kappa_2 & 0 & 0 & 0 & 0 & \kappa_1 & 0  \\
   0 & -\kappa_1 & 0 & 0 & (\Delta_1^{\Xi}+\Delta_2^{\Xi}) & 0 & {\kappa_2 } & 0  \\
   \kappa_1 & 0 & 0 & -(\Delta_1^{\Xi}+\Delta_2^{\Xi}) & 0 & -\kappa_2 & 0 & 0  \\
   0 & 0 & 0 & 0 & \kappa_2 & 0 & \Delta_1^{\Xi} & 0  \\
   0 & 0 & -\kappa_1 & -\kappa_2 & 0 & -\Delta_1^{\Xi} & 0 & \sqrt{3}\kappa_1  \\
   0 & 0 & 0 & 0 & 0 & 0 & -\sqrt{3}\kappa_1 & 0  \\
\end{array}} \right].\eqno (24)$$
Similarly for Eq.(17), the matrix of the lambda system is given by
$$ M_{ij}^{\Lambda} =
\left[ {\begin{array}{*{20}c}
   0 & \Delta_2^{\Lambda} & 0 & 0  & 0 & 0 & {-\kappa_1} & 0  \\
   -{\Delta_2^{\Lambda} } & 0 & 2\kappa_2 & 0 & 0 & {\kappa_1 } & 0 & 0  \\
   0 & {-2\kappa_2 } & 0 & 0 & {-\kappa_1 } & 0 & 0 & 0  \\
   0 & 0 & 0 & 0 & \Delta_1^{\Lambda} & 0 & {\kappa_2 } & 0  \\
   0 & 0 & {\kappa_1 } & -\Delta_1^{\Lambda} & 0 & {-\kappa_2 } &0 & {\sqrt 3 \kappa_1 }  \\
   0 & {\kappa_1} & 0 & 0 & {\kappa_2 } & 0 & (\Delta_1^{\Lambda}-\Delta_2^{\Lambda}) & 0  \\
   {\kappa_1} & 0 & 0 & {-\kappa_2} & 0 & -(\Delta_1^{\Lambda}-\Delta_2^{\Lambda}) & 0 & 0  \\
   0 & 0 & 0 & 0 & {-\sqrt 3\kappa_1 } & 0 & 0 & 0  \\
\end{array}} \right],\eqno (25)$$\\
and finally from Eq.(19) for the vee system we have,
$$ M_{ij}^{V} =
\left[ {\begin{array}{*{20}c}
   0 & (\Delta_1^{V}-\Delta_2^{V}) & 0 & 0  & {-\kappa_2} & 0 & {-\kappa_1} & 0  \\
   -(\Delta_1^{V}-\Delta_2^{V}) & 0 & 0 & {-\kappa_2} & 0 & {-\kappa_1 } & 0 & 0  \\
   0 & 0 & 0 & 0 & {-\kappa_1 } & 0 & {\kappa_2} & 0  \\
   0 & {-\kappa_2} & 0 & 0 & \Delta_1^{V} & 0 & 0 & 0  \\
   {\kappa_2} & 0 & {\kappa_1 } & -\Delta_1^{V} & 0 & 0 & 0 & {\sqrt 3 \kappa_1 }  \\
   0 & {\kappa_1 } & 0 & 0 & 0 & 0 & -\Delta_2^{V} & 0  \\
   {\kappa_1} & 0 & {-\kappa_2} & 0 & 0 & -\Delta_2^{V} & 0 & {\sqrt 3 \kappa_2}  \\
   0 & 0 & 0 & 0 & {-\sqrt 3 \kappa_1 } & 0 & {-\sqrt 3 \kappa_2} & 0  \\
\end{array}} \right],\eqno (26)$$\\
respectively.
\par
The algebraic structure of the $SU(3)$ group allows the existence of a set of quadratic casimirs which will appear in form of the quadratic constants. However,
the searching of the exact tuple of the Bloch vectors forming such constants is quite cumbersome because of the large number of such combinations. More specifically, for eight Bloch vectors, we have to search $\frac{8!}{(8-n)!n!} (=8, 28, 56, 70,...,1)$ number of combinations forming such tuples for $n=1,2,3,..,8$, respectively. We have developed a {\it Mathematica} program to search the exact combination of the Bloch vectors obtained by solving the Bloch equation in Eq.(23) and after a rigorous search obtain the following results for $n=3$ and $n=5$ only:
\par
For the cascade system at resonance ($\Delta_1^\Xi=0=\Delta_2^\Xi$), the Bloch space is split into two parts; one being of two sphere ${\mathcal S}^2$,
$$ {S_1^{\Xi}}^2(t)+{S_5^{\Xi}}^2(t)+{S_6^{\Xi}}^2(t)=
{S_1^{\Xi}}^2 (0)+{S_5^{\Xi}}^2 (0)+{S_6^{\Xi}}^2 (0),\eqno (27a)$$
and other is the four sphere ${\mathcal S}^4$,
$${S_2^{\Xi}}^2(t)+{S_3^{\Xi}}^2(t)+{S_4^{\Xi}}^2(t)+{S_7^{\Xi}}^2(t)+{S_8^{\Xi}}^2(t)=$$
$${S_2^{\Xi}}^2 (0)+{S_3^{\Xi}}^2 (0)+{S_4^{\Xi}}^2 (0)+{S_7^{\Xi}}^2 (0)+{S_8^{\Xi}}^2 (0),\eqno (27b)$$
respectively, where ${S_i^{\Xi}}(0)$ be the constants at $t=0$ which are to be evaluated in terms of the probability amplitudes. By noting the fact that the density matrix can be written as $\rho^{\Xi}(t)=U^{\dag}(t)\rho^{\Xi}(0)U(t)$, Eq.(20) becomes,
$$S_{i}^{\Xi}(0)=Tr[\rho^{\Xi}(0)\lambda_i]. \eqno (28)$$
Plucking back Eqs.(5) and (12) into Eq.(28), different values of the constants $S_a^\Xi(0)$ can be evaluated in terms of probability amplitudes ${C^\Xi_{1,2,3}}(0)$ at $t=0$ and Eqs.(27) become
$${S_1^{\Xi}}^2+{S_5^{\Xi}}^2+{S_6^{\Xi}}^2=4{|C_1^{\Xi}}(0)|^2{|C_2^{\Xi}}(0)|^2+4{|C_2^{\Xi}}(0)|^2{|C_3^{\Xi}}(0)|^2 \eqno (29a)$$
and
$$ {S_2^{\Xi}}^2+{S_3^{\Xi}}^2+{S_4^{\Xi}}^2+{S_7^{\Xi}}^2+{S_8^{\Xi}}^2=
\frac{4}{3}({|C_1^{\Xi}}(0)|^2+{|C_2^{\Xi}}(0)|^2+{|C_3^{\Xi}}(0)|^2)^2$$
$$-3|{C_1}^{\Xi}(0)|^2|{C_2}^{\Xi}(0)|^2-3|{C_2}^{\Xi}(0)|^2|{C_3}^{\Xi}(0)|^2, \eqno (29b)$$\\
respectively. Similarly at resonance ($\Delta_1^\Lambda=0=\Delta_2^\Lambda$), for the lambda system we have two Bloch spheres,
$${S_1^{\Lambda}}^2+{S_4^{\Lambda}}^2+{S_7^{\Lambda}}^2=4{|C_1^{\Xi}}(0)|^2{|C_2^{\Xi}}(0)|^2+4{|C_3^{\Xi}}(0)|^2{|C_1^{\Xi}}(0)|^2 \eqno (30a)$$
and
$${S_2^{\Lambda}}^2+{S_3^{\Lambda}}^2+{S_5^{\Lambda}}^2+{S_6^{\Lambda}}^2+{S_8^{\Lambda}}^2=
\frac{4}{3}({|C_1^{\Lambda}}(0)|^2+{|C_2^{\Lambda}}(0)|^2+{|C_3^{\Lambda}}(0)|^2)^2$$
$$-3|{C_1}^{\Lambda}(0)|^2|{C_2}^{\Lambda}(0)|^2-3|{C_3}^{\Lambda}(0)|^2|{C_1}^{\Lambda}(0)|^2. \eqno (30b)$$\\
Also for the vee system at resonance ($\Delta^V_1=0=\Delta^V_2$), we have
$${S_2^{V}}^2+{S_4^{V}}^2+{S_6^{V}}^2=4{|C_2^{V}}(0)|^2{|C_3^{V}}(0)|^2+4{|C_3^{V}}(0)|^2{|C_1^{V}}(0)|^2 \eqno (31a)$$
and
$${S_1^{V}}^2+{S_3^{V}}^2+{S_5^{V}}^2+{S_7^{V}}^2+{S_8^{V}}^2=
\frac{4}{3}({|C_1^{V}}(0)|^2+{|C_2^{V}}(0)|^2+{|C_3^{V}}(0)|^2)^2 $$
$$-3|{C_2}^{V}(0)|^2|{C_3}^{V}(0)|^2-3|{C_3}^{V}(0)|^2|{C_1}^{V}(0)|^2,  \eqno (31b)$$\\
respectively. Thus we note that the 3- and 5-tuple of the Bloch vectors form two quadratic constants which are different for different three-level systems. From Eqs.(29-31) it is further evident that {\it at resonance}, the Bloch space ${\mathcal S}^7$ is broken into two-subspaces ${\mathcal S}^2\times{\mathcal S}^4$, which indicates that each configuration has distinct norm. It is instructive to note that at off resonance ($\Delta_1^a\neq0\neq\Delta_2^a$), the solutions of the Bloch equation of all three configurations satisfy,
$${S_1^{a}}^2+{S_2^{a}}^2+{S_3^{a}}^2+{S_4^{a}}^2+{S_5^{a}}^2+{S_6^{a}}^2+
{S_7^{a}}^2+{S_8^{a}}^2=
\frac{4}{3}({|C_1^{a}}(0)|^2+{|C_2^{a}}(0)|^2+{|C_3^{a}}(0)|^2)^2.\eqno (32)$$\\
This shows that the normalization condition ${|C_1^{a}}(0)|^2+{|C_2^{a}}(0)|^2+{|C_3^{a}}(0)|^2=1$ gives that the radius-squared of the seven-dimensional Bloch sphere ${\mathcal S}^{7}$ will be $\frac{4}{3}$ regardless of the configurations.
\par
Finally we consider the wave function of the qutrit system. A convenient parametrization of the normalized qutrit wave function is
$$|q_T>=\cos\frac{\theta_0}{2}|0>+\sin\frac{\theta_0}{2} \sin\frac{\theta_1}{2} \sin\frac{\theta_2}{2} e^{i\phi}|1>+$$
$$(\sin\frac{\theta_0}{2} \cos\frac{\theta_1}{2}+ i \sin\frac{\theta_0}{2} \sin\frac{\theta_1}{2} \cos\frac{\theta _2}{2})|2 >,\eqno(33)$$ \\
where the three coordinates $\theta_0$, $\theta_1$ and $\theta_2$ lie between $0$ to $\pi$ with arbitrary value of the phase angle $\phi$. From Eq.(33) it is evident that, the pure qutrit states $|0>$, $|1>$ and $|2>$ can be obtained by choosing the coordinates $(\theta_0,\theta_1,\theta_2)$ to be $(0,\theta_1,\theta_2)$, $(\pi,\pi,\pi)$ and $(\pi,0,\theta_2)$, respectively. This indicates that for a qutrit system, the state $|0>$ may be any point on the $(\theta_1,\theta_2)$ plane, $|1>$ corresponds to a well-defined point and $|3>$ lies on a line for any value of $\theta_3$, respectively. In other words, in the three-level scenario, the degeneracy of the $|0>$ and $|2>$ states shows that the Bloch space structure of the qutrit is highly non-trivial.
This is in contrast with the two-level system representing qubit, where the basis states $|0>$ and $|1>$ correspond to the north and south poles of the Bloch sphere, respectively [3,9]. Using Eq.(12), various elements of the density matrix corresponding to the qutrit wave function in Eq.(33) are given by
$$\rho_T^{11}=\cos^2\frac{\theta_1}{2}$$
$$\rho_T^{22}=\sin^2\frac{\theta_0}{2}\sin^2\frac{\theta_1}{2}\sin^2\frac{\theta_2}{2}$$
$$\rho_T^{33}=\frac{1}{4}(3+\cos\theta_1+\cos\theta_2-\cos\theta_1\cos\theta_2)\sin^2\frac{\theta_0}{2}$$
$$\rho_T^{12}={\rho_T^{21}}^{*}=\frac{1}{2}e^{i\phi}\sin\theta_0\sin\frac{\theta_1}{2}\sin\frac{\theta_2}{2}$$
$$\rho_T^{23}={\rho_T^{32}}^{*}=e^{-i\phi}\sin^2\frac{\theta_0}{2}\sin\frac{\theta_1}{2}(\cos\frac{\theta_1}{2}+i\sin\frac{\theta_1}{2}\cos\frac{\theta_2}{2})\sin\frac{\theta_2}{2}$$
$$\rho_T^{13}={\rho_T^{31}}^{*}=\frac{1}{2}\sin\theta_0(\cos\frac{\theta_1}{2} + i \sin\frac{\theta_1}{2}\cos\frac{\theta_2}{2})\eqno (34)$$
It is worth noting that this qutrit wave function is normalized, i.e., $Tr[\rho_T]=1$ and the density matrix satisfies the pure state condition, namely, $\rho_T^2=\rho_T$. Finally plucking back $\rho_T$ in Eq.(20) we obtain the Bloch sphere on $S^7$,
$${S_1^{a}}^2+{S_2^{a}}^2+{S_3^{a}}^2+{S_4^{a}}^2+{S_5^{a}}^2+{S_6^{a}}^2+
{S_7^{a}}^2+{S_8^{a}}^2=
\frac{4}{3}. \eqno (35)$$\\
This is precisely same as Eq.(32) obtained from different configurations of three-level system indicating the equivalence between the qutrit system with the three-level system.
\par
In this paper we have discussed different semiclassical three-level configurations and obtain the solution of corresponding Bloch equations. It is shown that, if the energies of the three levels follow a definite hierarchy prescription, it is possible to write the Hamiltonians in the $SU(3)$ basis which are different for different configurations. At zero detuning, the Bloch sphere ${\mathcal S}^7$ of each configuration is broken into two disjoint sectors ${\mathcal S}^2{\times}{\mathcal S}^4$ due to the existence of different constraints which are not alike. A significant outcome of our approach is that we have given a possible representation of the qutrit wave function which can be identically represented by a three-level system. Although our formulation gives a possible representation of the qutrit wave function, but it is quite difficult to ascertain the complicated structure of the Bloch space at this stage. Apart from that, the mechanism of minimizing the information loss while teleporting between two or multiple number of qutrits, developing appropriate search algorithm, the qutrit based protocol transfer etc are still open issues which require further exploration of the qutrit wave function proposed here. Although the quantum information processing has made significant progress with the conventional qubit as the primitive identity, but the potential application of the qutrit can not be properly addressed unless we do not understand the wave function of the quatrit and its relation with the Bloch space structure of the three-level system in greater detail. More exploration along this direction may provide some interesting results which are within reach of the future high-Q cavity as well as nanomaterial based experiments.
\vfill
\begin{center}
\large {\bf Acknowledgement}
\end{center}
SS is thankful to Department of Science and Technology, New Delhi for partial support and to S N Bose National Centre for Basic Sciences, Kolkata, for supporting his visit to the centre through the `Associateship' program.
\pagebreak
\bibliographystyle{plain}

\pagebreak

\begin{figure}
\begin{center}\begin{picture}(300,86)(0,0)

\Line(100,90)(200,90)

\Text(230,90)[]{$|3>$}

\Text(70,90)[]{$E_3$}

\ArrowLine(155,50)(155,85)

\Text(140,70)[]{$\Omega_2$}

\Line(100,50)(200,50)

\Text(70,50)[]{$E_2$}

\Text(230,50)[]{$|2>$}

\DashLine(100,85)(200,85){5}

\DashLine(100,45)(200,45){5}

\ArrowLine(145,10)(145,45)

\Text(130,30)[]{$\Omega_1$}

\Line(100,10)(200,10)

\Text(70,10)[]{$E_1$}

\Text(230,10)[]{$|1>$}

\LongArrow(110,98)(110,90) \LongArrow(110,78)(110,85)

\LongArrow(195,58)(195,50) \LongArrow(195,38)(195,45)

\Text(90,88)[]{$\Delta_1$}

\Text(210,48)[]{$\Delta_2$}

\end{picture} \\
\end{center}
\vspace{1.5cm}
\begin{center}\begin{picture}(300,86)(0,0)

\Line(100,90)(200,90)

\Text(210,89)[]{$\Delta_2$}

\Text(230,90)[]{$|2>$}

\Text(70,90)[]{$E_2$}

\ArrowLine(155,85)(170,40)

\Text(175,70)[]{$\Omega_2$}

\Line(100,40)(200,40)

\Text(230,40)[]{$|3>$}

\Text(70,40)[]{$E_3$}

\DashLine(100,80)(175,80){5} \DashLine(150,85)(200,85){5}

\LongArrow(110,100)(110,90) \LongArrow(110,70)(110,80)

\LongArrow(195,100)(195,90) \LongArrow(195,75)(195,85)

\ArrowLine(120,10)(150,80) \Text(125,50)[]{$\Omega_1$}

\Line(100,10)(200,10)

\Text(230,10)[]{$|1>$}

\Text(70,10)[]{$E_1$}

\Text(90,85)[]{$\Delta_1$}

\end{picture} \\
\end{center}
\vspace{1.5cm}
\begin{center}\begin{picture}(300,86)(0,0)

\Line(100,90)(200,90)

\Text(230,90)[]{$|1>$}

\Text(70,90)[]{$E_1$}

\ArrowLine(120,85)(140,10)

\Text(165,20)[]{$\Omega_2$}

\Line(100,40)(200,40)

\Text(230,40)[]{$|3>$}

\Text(70,40)[]{$E_3$}

\DashLine(100,85)(150,85){5}

\DashLine(140,35)(200,35){5}

\ArrowLine(145,10)(160,35)

\Text(120,60)[]{$\Omega_1$}

\Line(100,10)(200,10)

\Text(230,10)[]{$|2>$}

\Text(70,10)[]{$E_2$}

\LongArrow(110,98)(110,90) \LongArrow(110,78)(110,85)

\LongArrow(195,48)(195,40) \LongArrow(195,28)(195,35)

\Text(90,88)[]{$\Delta_1$}

\Text(210,38)[]{$\Delta_2$}

\end{picture} \\
\end{center}
\caption{\small{\bf Cascade, lambda and vee type three-level configurations
according to the Eberly-Hioe scheme [29] with the energy conditions, $E_3>E_2>E_1$,
$E_2>E_3>E_1$ and $E_1>E_3>E_2$, respectively. Here the position of the middle level (level-2)
with energy $E_2$ is changed to generate all models.}}
\end{figure}
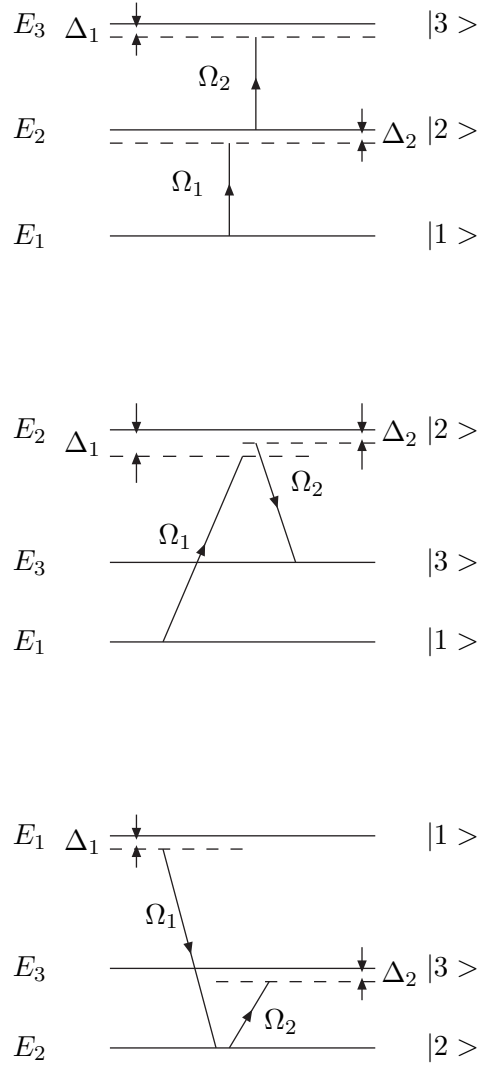


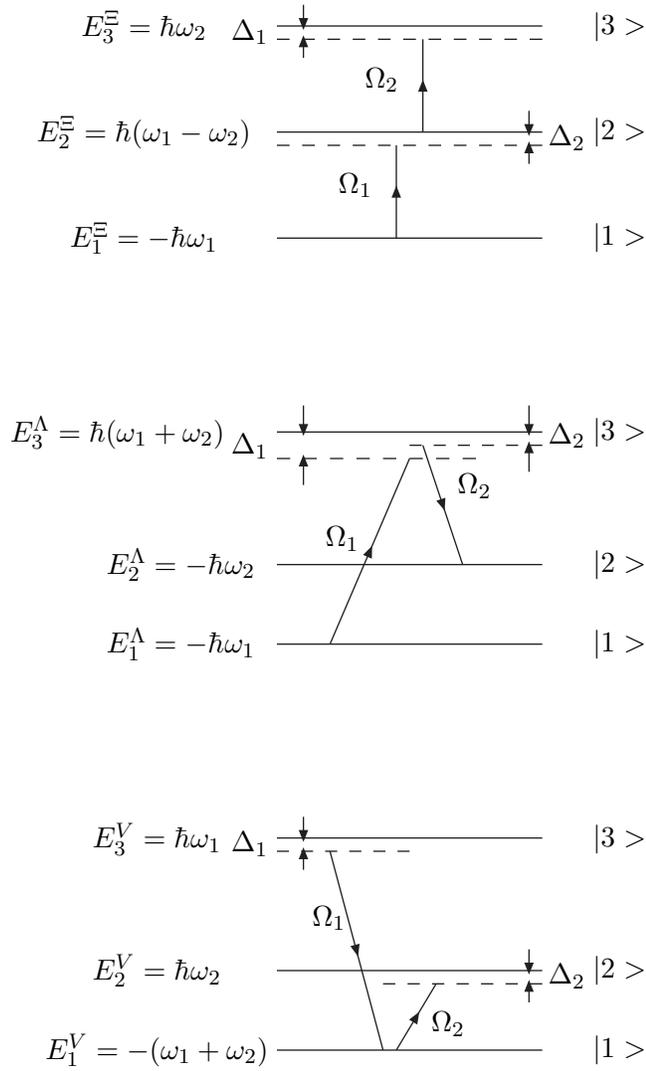
\begin{figure}
\begin{center}\begin{picture}(300,86)(0,0)

\Line(100,90)(200,90)

\Text(230,90)[]{$|3>$}

\Text(50,90)[]{$E_3^\Xi=\hbar\omega_2$}

\ArrowLine(155,50)(155,85)

\Text(140,70)[]{$\Omega_2$}

\Line(100,50)(200,50)

\Text(50,50)[]{$E_2^\Xi=\hbar(\omega_1-\omega_2)$}

\Text(230,50)[]{$|2>$}

\DashLine(100,85)(200,85){5}

\DashLine(100,45)(200,45){5}

\ArrowLine(145,10)(145,45)

\Text(130,30)[]{$\Omega_1$}

\Line(100,10)(200,10)

\Text(50,10)[]{$E_1^\Xi=-\hbar\omega_1$}

\Text(230,10)[]{$|1>$}

\LongArrow(110,98)(110,90) \LongArrow(110,78)(110,85)

\LongArrow(195,58)(195,50) \LongArrow(195,38)(195,45)

\Text(90,88)[]{$\Delta_1$}

\Text(210,48)[]{$\Delta_2$}

\end{picture} \\
\end{center}
\vspace{1.5cm}
\begin{center}\begin{picture}(300,86)(0,0)

\Line(100,90)(200,90)

\Text(210,89)[]{$\Delta_2$}

\Text(230,90)[]{$|3>$}

\Text(40,90)[]{$E_3^\Lambda=\hbar(\omega_1+\omega_2)$}

\ArrowLine(155,85)(170,40)

\Text(175,70)[]{$\Omega_2$}

\Line(100,40)(200,40)

\Text(230,40)[]{$|2>$}

\Text(64,40)[]{$E_2^\Lambda=-\hbar\omega_2$}

\DashLine(100,80)(175,80){5} \DashLine(150,85)(200,85){5}

\LongArrow(110,100)(110,90) \LongArrow(110,70)(110,80)

\LongArrow(195,100)(195,90) \LongArrow(195,75)(195,85)

\ArrowLine(120,10)(150,80) \Text(125,50)[]{$\Omega_1$}

\Line(100,10)(200,10)

\Text(230,10)[]{$|1>$}

\Text(64,10)[]{$E_1^\Lambda=-\hbar\omega_1$}

\Text(90,85)[]{$\Delta_1$}

\end{picture} \\
\end{center}
\vspace{1.5cm}
\begin{center}\begin{picture}(300,86)(0,0)

\Line(100,90)(200,90)

\Text(230,90)[]{$|3>$}

\Text(55,90)[]{$E_3^V=\hbar\omega_1$}

\ArrowLine(120,85)(140,10)

\Text(165,20)[]{$\Omega_2$}

\Line(100,40)(200,40)

\Text(230,40)[]{$|2>$}

\Text(55,40)[]{$E_2^V=\hbar\omega_2$}

\DashLine(100,85)(150,85){5}

\DashLine(140,35)(200,35){5}

\ArrowLine(145,10)(160,35)

\Text(120,60)[]{$\Omega_1$}

\Line(100,10)(200,10)

\Text(230,10)[]{$|1>$}

\Text(55,10)[]{$E_1^V=-(\omega_1+\omega_2)$}

\LongArrow(110,98)(110,90) \LongArrow(110,78)(110,85)

\LongArrow(195,48)(195,40) \LongArrow(195,28)(195,35)

\Text(90,88)[]{$\Delta_1$}

\Text(210,38)[]{$\Delta_2$}
\end{picture} \\
\end{center}
\caption{\small{\bf Cascade, lambda and vee  type three-level configurations according to our scheme [45] $E_3^a>E_2^a>E_1^a$ ($a=\Xi, \Lambda$ and $V$).}}
\end{figure}

\end{document}